# Exploring First Stars Era with GLAST

A. Kashlinsky[1,2] and D. Band[1,3,4]

[1] *Observational Cosmology Lab, Goddard Space Flight Center, Greenbelt MD 20771,* [2] *SSAI,* [3] *UMBC,* [4] *CRESST*

**Abstract.** Cosmic infrared background (CIB) includes emissions from objects inaccessible to current telescopic studies, such as the putative Population III, the first stars. Recently, strong direct evidence for significant CIB levels produced by the first stars came from CIB fluctuations discovered in deep Spitzer images. Such CIB levels should have left a unique absorption feature in the spectra of high-z GRBs and blazars as suggested in [4]. This is observable with GLAST sources at z>2 and measuring this absorption will give important information on energetics and constituents of the first stars era.

Cosmic infrared background (CIB) is a repository of emission throughout the entire history of the Universe, including from epochs containing objects inaccessible to current telescopic studies (see [1] for review). One such epoch is when the first stars are thought to have been formed corresponding to $z>10$. If the first stars (commonly called Population III – Pop III) were massive, they should have produced significant near-IR (NIR) CIB which should be cut-off by Lyman absorption at $< 1$ μm [2]. CIB derived from DIRBE- and IRTS-based analyses suggest higher fluxes at ~1-5 μm than that obtained by integrating galaxy counts [1]. The net CIB flux produced by these observed populations saturates at $m_{AB}$~20-21 with fainter galaxies contributing little to the total budget. The excess flux at the near-IR (NIR) is commonly referred to as the NIRBE. The NIRBE, if extragalactic, must thus originate in still fainter systems, likely located at very early cosmic times. It is important to emphasize that e.g. K-band galaxies at $m_{AB}$~19 are only at z~1 (e.g. [3]) with ordinary galaxies at earlier times contributing very little CIB; any modeling must account for the CIB evolution as observed in galaxy counts. The entire NIRBE correspond ~30 nW/m²/sr at 1-4 μm of and this can be accounted for if only ~2% of the baryons have been processed through Pop III by z~10 [4].

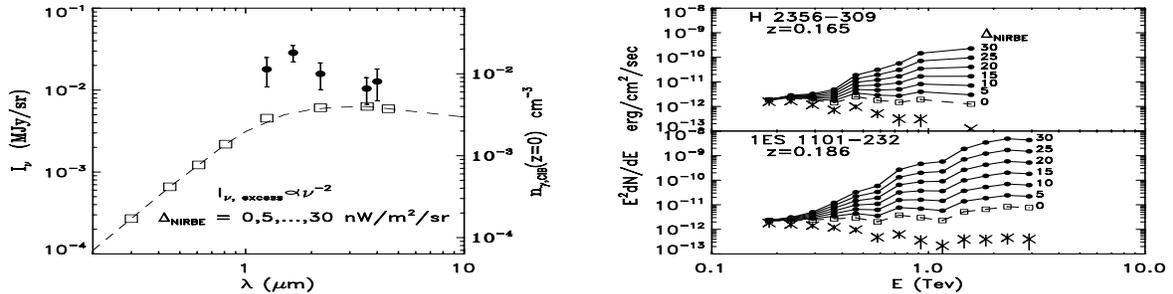

**FIGURE 1.** Right: Squares show net CIB in MJy/sr from integrating galaxy counts and the dashed line corresponds to the CIB in the absence of NIRBE. Comoving number density of these photons is also shown. Circles show the NIRBE levels as discussed in [1]. Left: HESS measured spectra for the two blazars [5] are shown with crosses. Circles show the spectra corrected for absorption due to CIB photons produced by Pop III at z>10 and with total amplitude normalized to the shown value of $\Delta_{NIRBE}$.

Recent HESS observations of absorption in the spectra of two z~0.2 blazars were taken to suggest that most of the NIRBE is not extragalactic for otherwise the unabsorbed spectra of the blazars would have to be too steep [5]. The situation is not as straightforward because the analysis in [5] used a template CIB which extends to <1 μm and is unsuitable for modeling CIB from z>7-10 sources, such as Pop III, whose CIB component, with a Lyman cutoff around 1 μm, should be used in any proper modeling. With this the constraints become significantly less severe and substantial CIB is still allowed by the data. This is shown in Fig.2, where we reconstructed the unabsorbed HESS blazar spectra using a more appropriate modeling of the CIB from Pop III. The model for the Pop III CIB component assumed: 1) CIB with $I_\nu \propto \nu^{-2}$, 2) CIB with a cutoff at the present wavelength of 1 μm, corresponding to the Lyman cutoff at z~10, 3) the amplitude normalized to the net NIRBE of amplitude $\Delta_{NIRBE}$ shown in the figure, and 4) the contribution from ordinary galaxies is given by the observed values derived from deep galaxy counts. The right panels show the uncorrected spectra for various levels of NIRBE. Using the limit on the hardness parameter ($\Gamma$>-1.5) from [5], at least half of the claimed NIRBE levels can still be produced by Pop III ($\Delta_{NIRBE}$<15 nW/m$^2$/sr).

The most direct evidence for substantial energy release during the first stars' era comes from the recent measurements of CIB fluctuations at 3.6-8 μm in deep Spitzer images, which remain after removing intervening ordinary galaxies to very faint levels [6,7]. The fluctuations are produced by populations that have significant clustering component (from clustering of the emitters), but only a small shot noise level (from sources occasionally entering the beam). The results indicate that 1) the CIB fluxes from cosmological populations producing these fluctuations are substantial with > 1-2 nW/m$^2$/sr at 3.6 and 4.5 μm, and 2) these CIB fluctuations are produced by sources with individual fluxes of only <10-20 nJy, which likely places them at z>10 [8].

If early stars produced even a fraction of the NIRBE, they would provide a source of abundant photons at high z. The present-day value of $I_\nu$=1 MJy/sr corresponds to the comoving number density of photons per logarithmic energy interval of $4\pi/c\, I_\nu h_{Planck}$=0.6 cm$^{-3}$; if these photons come from high z, their number density would increase as $(1+z)^3$ at early times. These photons would also have higher (blue-shifted) energies in the past and would thus provide abundance of absorbers for sources of sufficiently energetic photons at high z via two-photon absorption.

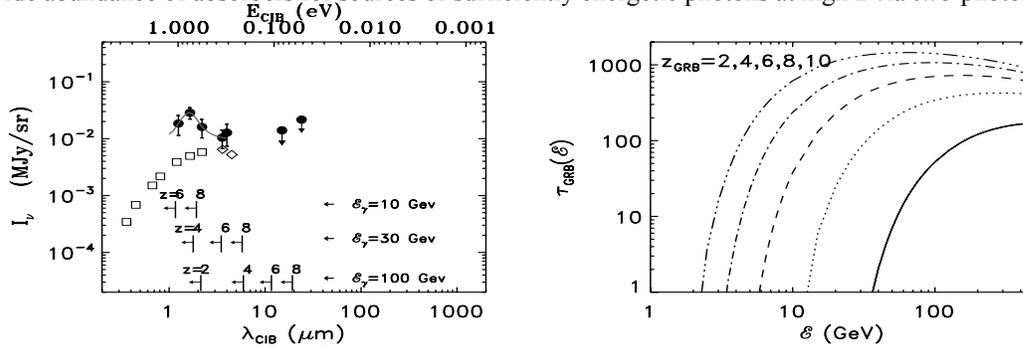

**FIGURE 2.** Left: shows the range of CIB photons which affects gamma-rays at given energy and the marked redshifts. Right: Optical depth vs γ-ray energy due to 2-photon absorption by the NIRBE photons for GRBs and other sources at marked redshifts.

Left panel of Fig.2 shows the range of the CIB wavelengths where the two-photon absorption operates for different z; GLAST/LAT observations of GRB's and blazars at z>1-2 should thus provide a test of emissions from the Pop III era. Fig. 2 shows the optical depth for sources at high z (e.g., GRBs) assuming the entire NIRBE originated at z>10 [4]. If so, then γ-rays with energies > 260(1+z)$^{-2}$ Gev from sources at z>1-2 should be completely absorbed even if only a faction of the NIRBE originated from the first stars era. GLAST and the Pop III era parameters have thus "conspired" very fortunately to uncover the CIB produced during the latter: 1) there should be a sharp cutoff at the Lyman limit at z>10, so CIB coming from these epochs is cut-off below the present-day wavelength of ~1μm; and 2) the threshold on the two-photon absorption process is such that with the GLAST energy limit of 300 GeV such measurement is not sensitive to the CIB beyond ~ 5 μm.

For our purposes GRBs are cosmological sources with smooth spectra that extend up to GeV energies and are ideal for detecting absorption in the intervening IGM. As discussed, a source at z>1-2 should have a spectral cutoff at $E_c$=260(1+z)$^{-2}$ GeV due to significantly energetic Pop III era emission. While GRBs are known to emit up to 18 GeV [9] and perhaps to TeV energies [10], the >100 MeV spectrum that the LAT will detect must be extrapolated from a few detections by the *CGRO*/EGRET. At the lower energies (10 keV to 1 MeV), where many bursts have been detected and characterized, the GRB spectrum can be described by a smoothly broken power law (the 'Band' function [11]); in 6 bursts the EGRET observations [12] are consistent with an extrapolation of the simultaneously observed lower energy emission with a high energy spectral index of β ~ -2 (where N(E)∝E$^\beta$). In a few bursts EGRET observed emission that lasted longer than the low energy emission (for GRB 940217 burst photons were observed up to 90 minutes after the 3 minute long lower energy emission [9]). Milagrito may have detected TeV emission from GRB 970717a [10]. In GRB 941017 EGRET detected an additional hard power law component simultaneous with the lower energy 'Band' function component [13]. Thus the EGRET observations demonstrate the existence of >1 GeV emission. In some cases the lower energy spectral components extrapolate to this energy band, but additional spectral and temporal components, expected on theoretical grounds [14,15] exist. Despite these uncertainties, we can estimate the number of LAT GRBs where absorption by Pop III era photons might be observed by extrapolating the <1 MeV (synchrotron) spectra up to >1 GeV for a GRB population based on the BATSE observations. The GRB rate as a function of intensity, the burst spectrum, and the LAT's effective area as a function of photon energy have been convolved to predict the LAT's burst detection rate (see [16] on the GRB rate per year).

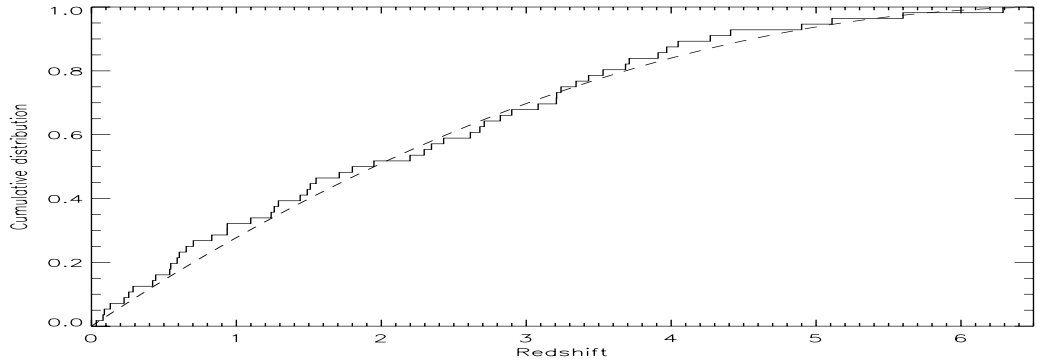

**FIGURE 3.** Distribution of z of *Swift*-detected GRBs (solid histogram) and the empirical function (dashes) used in calculations.

GRBs originate over a wide redshift range, and therefore we can expect the cutoff resulting from the Pop III era photon field to range over a wide energy range. As a result of a broad intrinsic energy distribution and possibly evolution, GRB fluence correlates poorly with z, and consequently GRBs that are bright in the LAT will not necessarily originate at lower z than dim LAT bursts. The z-distribution of the bursts the LAT will detect is currently uncertain: this distribution can be calculated assuming GRBs are correlated with other astrophysical phenomena (e.g., the star formation rate), or can be estimated empirically. We use the observed distribution from the *Swift* mission (Fig. 3), which detects and rapidly localizes ~100 GRBs per year [17]. By convolving the LAT detection rate of bursts in which $E_c$ can be observed, the relationship between $E_c$ and z, and the z-probability distribution, we estimate that the LAT will detect ~7 GRBs/year with observable Pop III era absorption cutoffs.

The stronger case for absorption by Pop III era photons will result from determining $E_c$ and z for the same burst. The fraction of LAT-observed bursts for which redshifts will be determined is uncertain. The *Swift* mission is projected to last well into the GLAST era, and *Swift* should observe ~1/6 of the bursts the LAT detects; currently redshifts have been determined for ~1/3 of the *Swift* bursts. GLAST will rapidly localize bursts onboard and on the ground, permitting follow-up ground-based observations that might result in burst redshifts; however, the localization by GLAST's GBM (degrees) and LAT (tenths of a degree) detectors may be too large to result in many redshifts. In addition, various relations between burst characteristics have been proposed that could turn bursts into 'standard candles.' For example, a tight relationship between the burst's peak luminosity, a measure of its duration, and the average photon energy has been found [18]. The two GLAST instruments will observe the burst lightcurve (from which the duration can be measured) and spectrum (from which the average photon energy can be measured). The predicted ratio between the observed peak flux and the peak luminosity gives an estimate of the redshift that is akin to a photometric redshift. Thus we estimate that enough bursts with LAT-determined cutoffs will have redshifts enabling robust determination of the Pop III era spectral absorption over the course of the GLAST mission.

AK acknowledges support from the National Science Foundations grant AST-0406587.